\documentstyle[twoside,fleqn,espcrc2,epsf]{article}
\newcommand{\lsim}{\raisebox{0.3mm}{\em $\, <$} 
\hspace{-3.3mm} \raisebox{-1.8mm}{\em $\sim \,$}}

\newcommand{\beq}{\begin{equation}}
\newcommand{\eeq}{\end{equation}}
\newcommand{\beqs}{\begin{eqnarray}}
\newcommand{\eeqs}{\end{eqnarray}}
\def\lsim{\raise0.3ex\hbox{$\;<$\kern-0.75em\raise-1.1ex
\hbox{$\sim\;$}}}
\def\gsim{\raise0.3ex\hbox{$\;>$\kern-0.75em\raise-1.1ex
\hbox{$\sim\;$}}}

\def\npb#1{Nucl.\ Phys.\ {\bf B\,#1}}

\def\plb#1{Phys.\ Lett.\ {\bf B\,#1}}
\def\prc#1{Phys.\ Rev.\ {\bf C\,#1}}
\def\prd#1{Phys.\ Rev.\ {\bf D\,#1}}
\def\prl#1{Phys.\ Rev.\ Lett. {\bf#1}}

\def\sjnp#1{Sov. J. Nucl. Phys. {\bf #1}}

\begin{document}
\topmargin = -1.2cm

\title{Status of the solutions to neutrino anomalies 
based on non-standard neutrino interactions
\thanks{Talk presented at ``NuFact'00'', Monterey, CA, USA,
May 22-26, 2000. 
}}

\author{Hiroshi Nunokawa\\
        {\ }\\
Instituto de F\' {\i}sica Gleb Wataghin,
Universidade Estadual de Campinas, UNICAMP\\    
13083-970 -- Campinas, Brazil.}

\begin{abstract}
We review the status of the solutions to neutrino anomalies by
flavor-changing as well as flavor-diagonal neutrino interactions.
While it is difficult to explain the atmospheric neutrino data 
the solar neutrino data can be well accounted for by the massless
neutrino oscillation induced by such non-standard neutrino 
interactions.
We also discuss the possibility to test such kind of 
interactions by the future neutrino oscillation experiments 
at neutrino factories. 
\end{abstract}

\maketitle

%%%%%%%%%%%%%%%%%%%%%%%%%%%%%%%%%%%%%%%%%%%%%%%%%%%%%%%%%%%%%%%%%%%%%%
%% Section I %%%%%%%%%%%%%%%%%%%%%%%%%%%%%%%%%%%%%%%%%%%%%%%%%%%%%%%%%
%%%%%%%%%%%%%%%%%%%%%%%%%%%%%%%%%%%%%%%%%%%%%%%%%%%%%%%%%%%%%%%%%%%%%%

\section{Introduction}

In this talk we review the present status of the 
solutions to the solar neutrino problem and 
atmospheric neutrino observations by the oscillation of
neutrinos which are massless (or degenerate in mass) 
induced by flavor-changing (FC) as well as flavor-diagonal (FD) 
interactions~\cite{wol,valle87,FY,GMP,ER,FCsols,BGHKN,FCatm} in matter. 
(The possibility to explain LSND signal by flavor-changing
interactions, but not due to the FC induced oscillation 
we will consider here, 
has been discussed in Ref.~\cite{BG}
where such possibility was discarded 
and we do not discuss it here.) 
We also consider the possibility to test such 
kind of interactions by future neutrino oscillation 
experiments at neutrino factories.

{}From a phenomenological point of view FC interactions 
of neutrinos alone (regardless of the presence of the 
FD interactions) can induce flavor conversion when neutrinos 
travel through matter~\cite{wol} even if neutrinos are massless.  
Moreover, the presence of FD interactions in addition to the 
FC one can induce a MSW-like resonant conversion of 
massless neutrinos in matter~\cite{valle87,FY,GMP}. 

Several models which can induce such massless 
neutrino conversion in matter exist. 
The simplest example of such mechanism was first 
considered in Ref.~\cite{valle87} where it is shown 
its possible implication for supernova neutrinos 
rather than solar and/or atmospheric neutrinos. 
The most plausible candidate of model which could be 
relevant for both solar and atmospheric neutrinos is 
the minimal supersymmetric standard model without 
$R$ parity~\cite{SUSYwoR} as it was first considered 
in Ref.~\cite{GMP,ER}.  
The other example (though less relevant for solar neutrinos)
is the model based on the extended gauge structure 
$SU(3)_C \otimes SU(3)_L \otimes U(1)_N$ (331 models)~\cite{331}. 

%For our phenomenological approach we simply assume 
Our approach here is completely phenomenological.  
We simply assume the existence of a tree-level process
$\nu_\alpha+ f \to \nu_\beta + f$ 
with an amplitude proportional to $\displaystyle
g_{\alpha f}g_{\beta f}/4m^2$ 
$(\equiv \epsilon_{\alpha\beta}\sqrt{2} G_F)$
where $\alpha$ and $\beta$ are flavor
indices, $f$ stands for the interacting elementary fermion (charged
lepton, $d$-like or $u$-like quark) and $g_{\alpha f}$ is the coupling
involved in the vertex where a $\nu_\alpha$ interacts with $f$ through
a scalar or vector boson of mass $m$. 
We try to fit the solar as well as atmospheric neutrino data 
treating $\epsilon_{\alpha\beta}$ as free parameters.

Secs. 2 and 3 are devoted for solar and atmospheric 
neutrinos, respectively. In sec. 4, we discuss 
the possibility to test this kind of interactions by
future oscillation experiments at neutrino factories. 
In sec. 5, we give conclusions. 

%%%%%%%%%%%%%%%%%%%%%%%%%%%%%%%%%%%%%%%%%%%%%%%%%%%%%%%%%%%%%%%%%%%%%%
%% Section II %%%%%%%%%%%%%%%%%%%%%%%%%%%%%%%%%%%%%%%%%%%%%%%%%%%%%%%%
%%%%%%%%%%%%%%%%%%%%%%%%%%%%%%%%%%%%%%%%%%%%%%%%%%%%%%%%%%%%%%%%%%%%%%

\section{Solar neutrino}
\vskip -0.1cm
The disagreement between the expected solar neutrino event rates
from theoretical predictions~\cite{BP98} and the observed ones
~\cite{solexp} is known as the solar neutrino problem~\cite{JNB}. 
While the most plausible solutions can be provided by 
the neutrino oscillation induced by mass and mixing
~\cite{vacuum,MSW} 
it has been proposed that massless neutrino conversion 
induced by non-standard interactions can also explain 
the observed deficit~\cite{GMP,FCsols}. 

Here, for simplicity, we consider the system of two neutrino 
flavor, $\nu_e -\nu_x (x=\mu,\tau$) which 
are massless or degenerate in mass. 
In the presence of FC and FD interactions 
in matter neutrino evolution equation is 
given as~\cite{GMP,FCsols}:
\vglue -0.3cm
\begin{eqnarray} &  i{\displaystyle{d}\over 
\displaystyle{dr}}\left( 
\begin{array}{c} \nu_e 
\\ \nu_x  \end{array} \right) = \hskip .3cm & \hskip-.1cm
H 
\left( \begin{array}{c} \nu_e  \\ \nu_x 
\end{array} \right) ,
\label{motion} 
\end{eqnarray}
where 
\begin{eqnarray} &  H 
 = \hskip .3cm & \hskip-.1cm
\sqrt{2}\,G_F \left( \begin{array}{cc} n_e(r) &  \epsilon n_f(r)
\\ \epsilon n_f(r)& \epsilon ' n_f(r) \end{array} \right)
\label{hamil} 
\end{eqnarray}
where, $\nu_a \equiv \nu_a (r)$, a=$e,\mu,\tau$ are the 
probability amplitudes to find these neutrinos at 
a distance $r$ from their creation position, 
$\sqrt{2}\,G_F n_f(r) \epsilon$ 
($\epsilon \equiv \epsilon_{e x }$) is the
$\nu_e + f \to \nu_x + f$ forward scattering amplitude and
$\sqrt{2}\,G_F n_f(r) \epsilon '$ 
($\epsilon' \equiv \epsilon_{xx} - \epsilon_{ee}$)
is the difference between the
$\nu_e - f$ and $\nu_x - f$ elastic forward scattering
amplitudes, with $n_f(r)$ being the number density of the fermions
which induce such processes.

%%%%%%%%%%%%%%%%%%%%%%%%%%%%%%%%%%%%%%%%%%%%%%%%%%%%%%%%%%%%%%%
\begin{figure}[h]
\vglue -0.8cm
\hbox to\hsize{\hss\epsfxsize=7.8cm\epsfysize=5.7cm\epsfbox{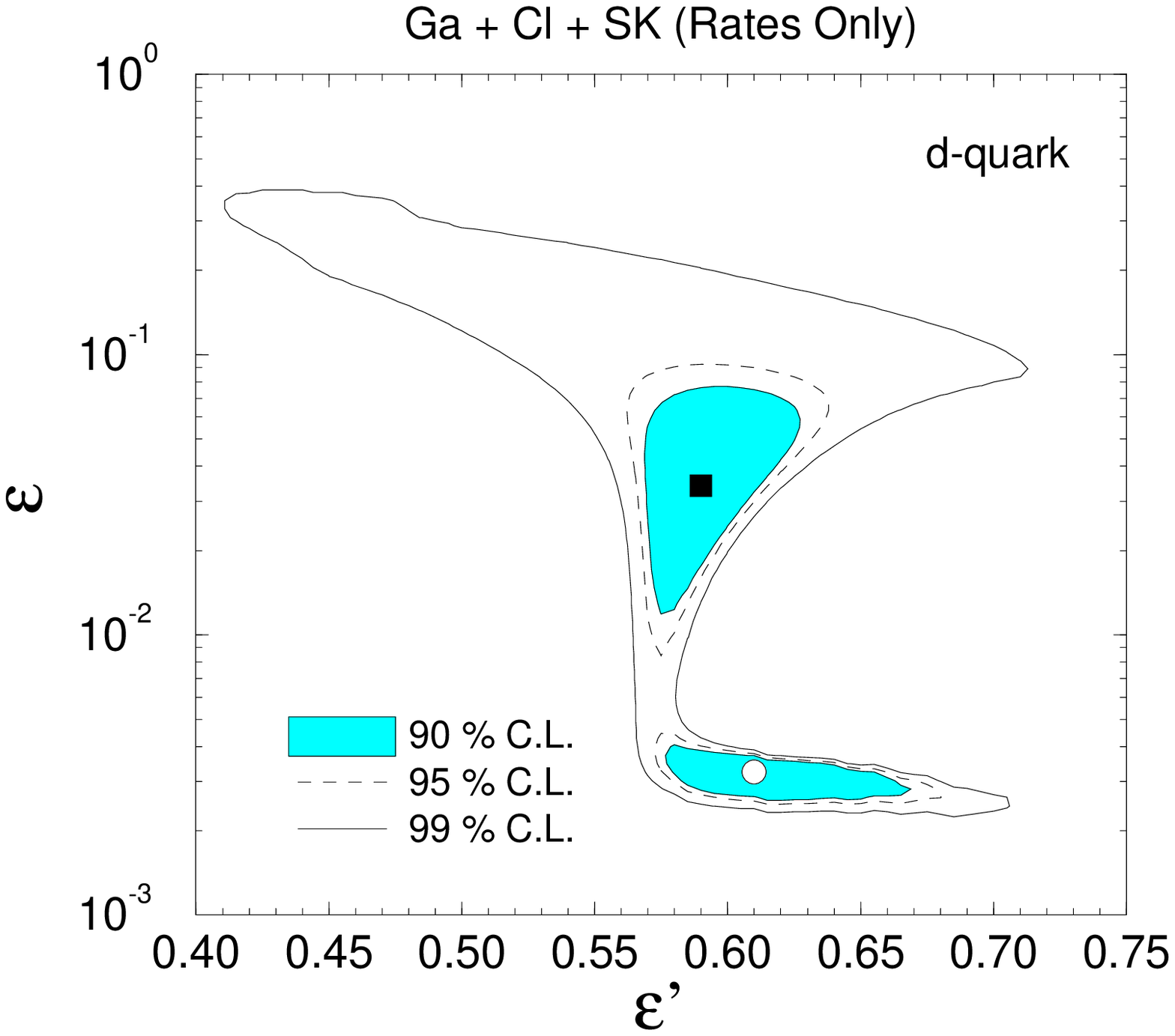}\hss}
\vglue -0.3cm
\hbox to\hsize{\hss\epsfxsize=7.8cm\epsfysize=5.7cm\epsfbox{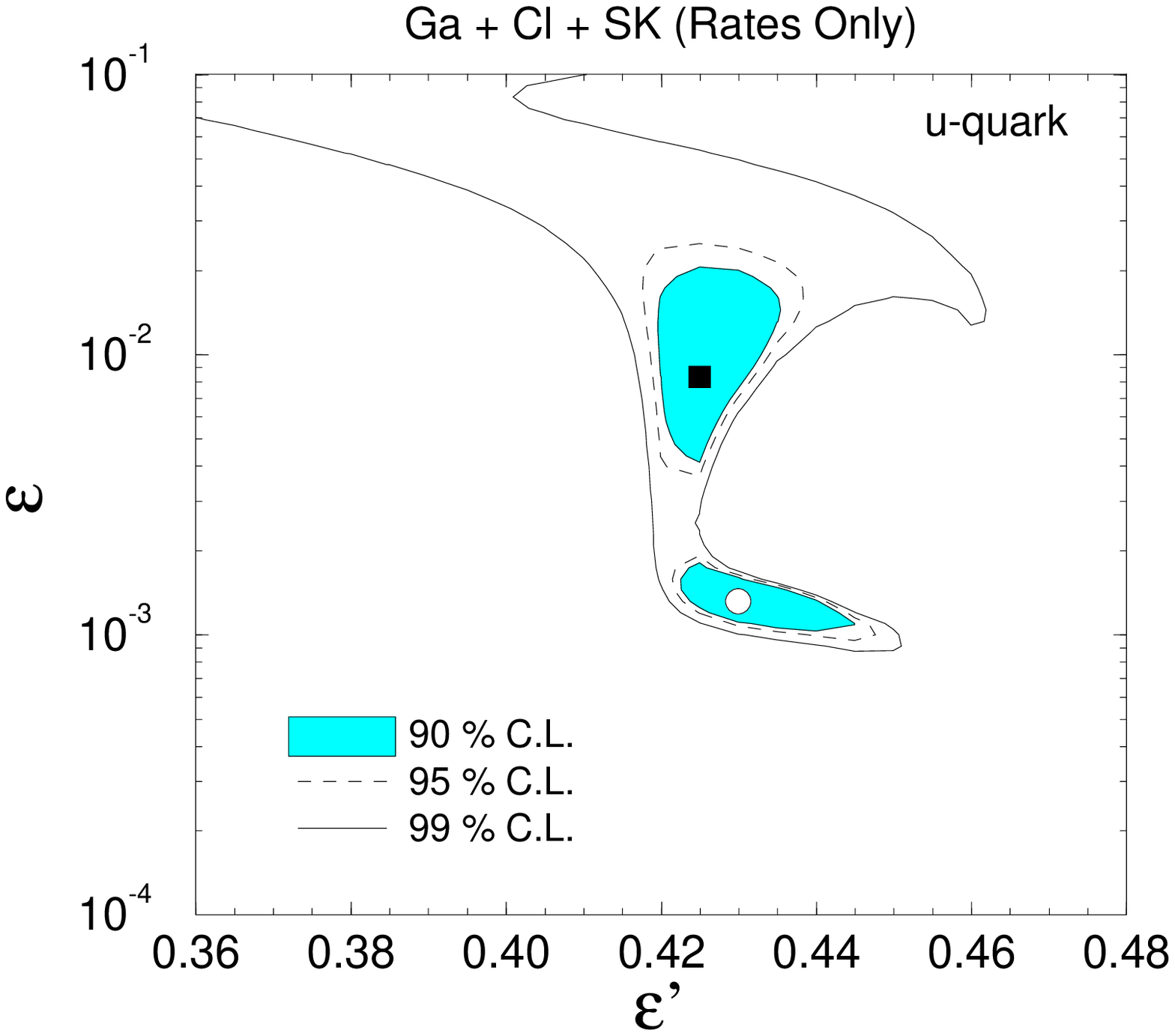}\hss}
\vglue -1.3cm
\label{fig:rates_fix_u} 
\caption{ {\small
Allowed region obtained by rates only for d-quark (upper panel) 
and u-quark (lower panel) interactions. 
The open circle (solid square) indicates 
the best fit (local best fit) point. 
(Taken from Ref.~\cite{BGHKN}.)} } 
\vglue -0.7cm
\end{figure}
%%%%%%%%%%%%%%%%%%%%%%%%%%%%%%%%%%%%%%%%%%%%%%%%%%%%%%%%%%%%%%%

A resonance conversion similar to MSW~\cite{MSW} can occur 
at some point where the condition 
\beq 
{\epsilon'} n_f(r_{res}) = n_e(r_{res}) \,, 
\label{res}
\eeq
as well as the corresponding adiabaticity 
condition are satisfied~\cite{MSW}. 
We note that such resonant flavor conversion can not occur 
for the case where the relevant fermion is electron ($f=e$) 
alone, and hence we do not consider this case here
(see, however, Ref.~\cite{fc_ele}). 

In this mechanism the conversion probability does
not depend on neutrino energy and hence one might
think it is impossible to explain the solar neutrino data
which indicate some energy dependent suppression
of neutrino fluxes. 
However, due to the difference in neutrino production 
distributions in the solar core, it is possible to 
suppress differently the neutrinos from difference
nuclear reactions in the solar core~\cite{GMP,FCsols,BGHKN}.  

\begin{figure}[h]
\vglue -0.6cm
\hbox to\hsize{\hss\epsfxsize=8.1cm\epsfysize=5.7cm\epsfbox{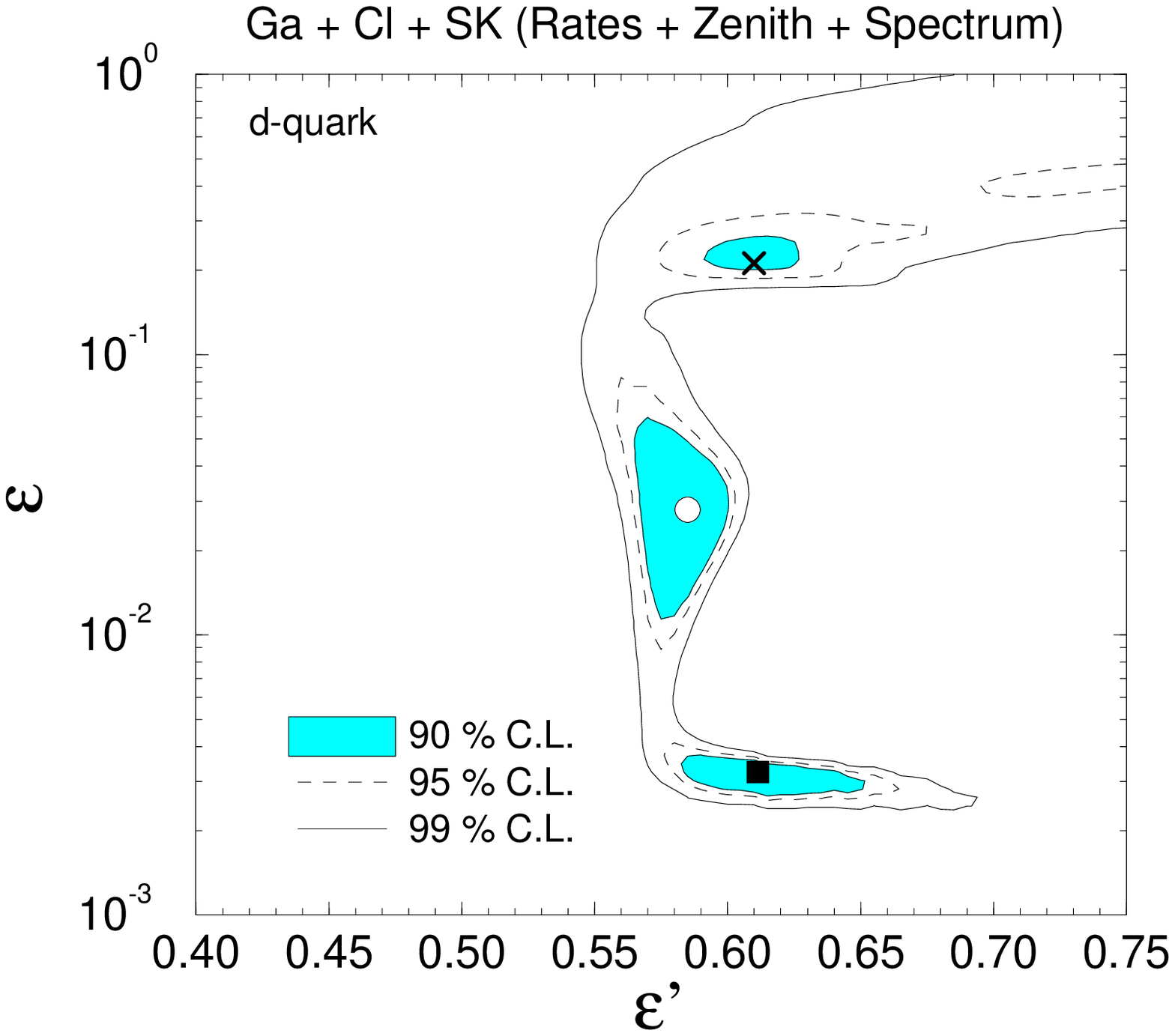}\hss}
\vglue -0.3cm
\hbox to\hsize{\hss\epsfxsize=7.4cm\epsfysize=5.7cm\epsfbox{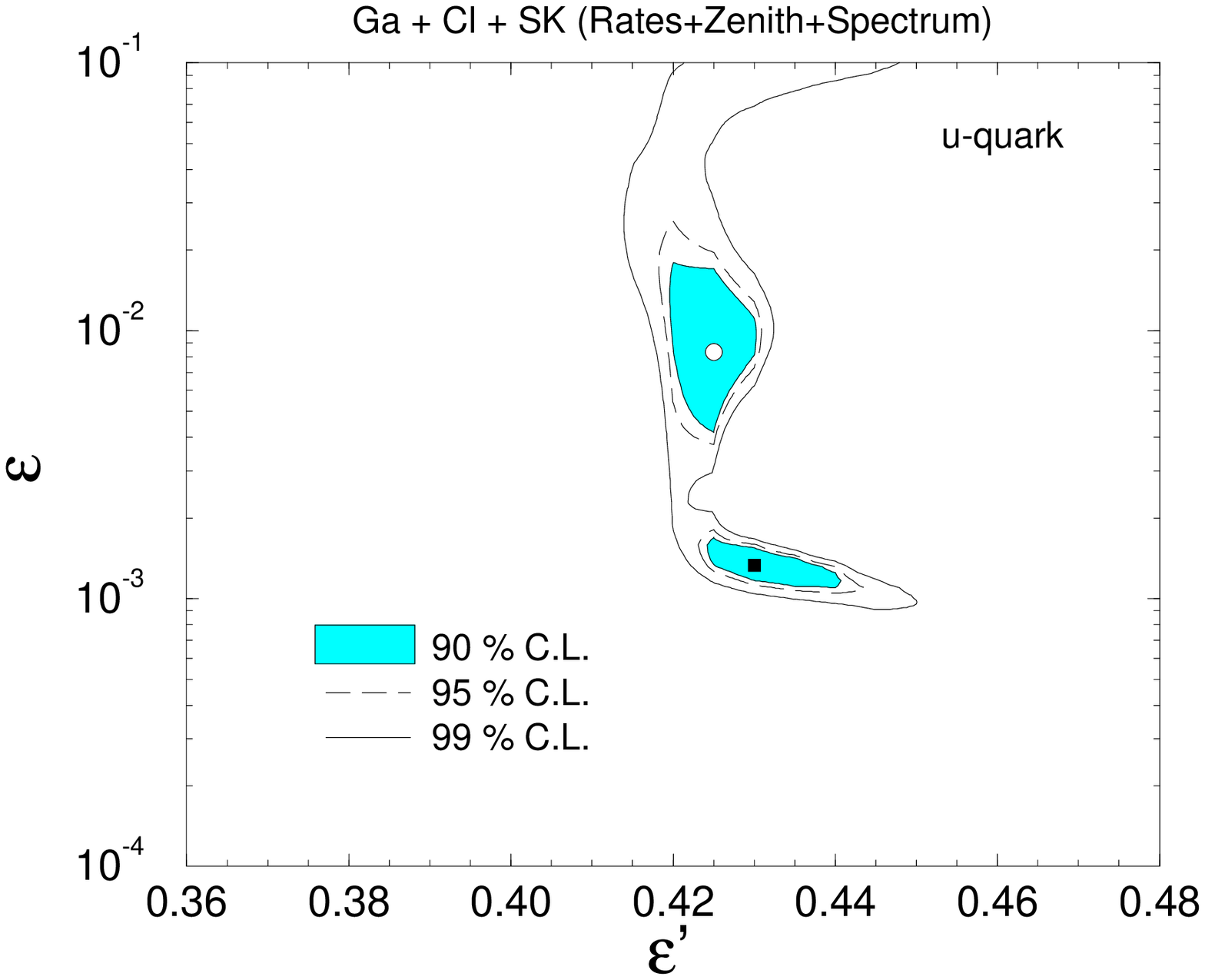}\hss}
\vglue -1.3cm
\caption{\label{fig:rates_zenith_fix_d}
{\small 
Same as in Fig. 1 but for the combined (rates+zenith+spectrum) 
analysis. The open circle (solid square and cross) indicates 
the best fit (local best fit) point. 
(Taken from Ref.~\cite{BGHKN}.)} } 
\vglue -0.6cm
\end{figure}

In Fig. 1 and 2 we present the region allowed by the total rates 
obtained by all the current solar neutrino experiments~\cite{solexp}
assuming the non-standard interactions 
with $d$-quark and $u$-quark in matter, respectively. 
In Fig. 3 and 4 we present the region allowed by the combined data
of the total rates, SK spectrum and SK zenith angle dependence. 
In Table I, we also present the values of $\chi^2_{min}$ for 
each case. From these results we conclude that the quality 
of the fit is quite good. 

Here let us briefly comment that $\epsilon$ and $\epsilon'$ parameters
can be constrained model independently %using the negative results 
due to the absence of the lepton flavor-violating process or 
lepton universality violation %decay searches 
as was discussed in Ref.~\cite{BGHKN}. 
It is found in Ref. \cite{BGHKN} that the magnitudes of  
$\epsilon$ and $\epsilon'$ parameters 
required to solve solar neutrino 
problem are still allowed for $\nu_e-\nu_\tau$ channel 
by the current experimental limits but not allowed for 
the $\nu_e-\nu_\mu$ one. This is because effects of such 
new physics (beyond the standard model) are much more 
constrained between the first and the second generations 
than that between the first and the third ones. 

%%%%%%%%%%%%%%  Table I  %%%%%%%%%%%%%%%%%%%%%%%%%%%%%%%%%%%%%%%%%%%%%%%%%
%
\begin{table}[h]
\vglue -0.3cm
\caption[Tab]
{ {\small
$\chi^2_{\rm{min}}$ values for our solar neutrino analysis. 
The numbers of degree of freedom are 2, 5 and 24 for 
rates, zenith and combined analysis, respectively. 
} }
\begin{center}
\begin{tabular}{c|c|c|c}
\hline
\hskip -0.5cm Case  & Rates &  Zenith  & Combined   \\ 
\hline
\hskip -0.5cm $d$-quark & 2.44    & 1.11   & 29.1    \\ 
\hskip -0.5cm $u$-quark & 2.75   & 1.44   & 28.5   \\ 
\hline
\end{tabular}
\end{center}
\label{tab1}
\vskip -1.15cm
\end{table}

More detailed discussions of our solar neutrino analysis 
can be found in Ref.~\cite{BGHKN}. 

%%%%%%%%%%%%%%%%%%%%%%%%%%%%%%%%%%%%%%%%%%%%%%%%%%%%%%%%%%%%%%%%%%%%%%
%% Section III %%%%%%%%%%%%%%%%%%%%%%%%%%%%%%%%%%%%%%%%%%%%%%%%%%%%%%%%
%%%%%%%%%%%%%%%%%%%%%%%%%%%%%%%%%%%%%%%%%%%%%%%%%%%%%%%%%%%%%%%%%%%%%%
\section{Atmospheric neutrino}

Atmospheric neutrino data obtained by several 
experiments~\cite{atmexp}, in particular  
the ones by the SuperKamiokande (SK)~\cite{skatm}
can be quite well accounted for by $\nu_\mu \rightarrow \nu_\tau$ 
mass induced neutrino oscillation and for this reason this 
was considered to be  a very  compelling evidence for non-zero 
neutrino masses~\cite{kajita98}.

Although such mass induced oscillation seems to be the most 
plausible explanation (see, for e.g., ~\cite{FGV} for 
a recent analysis), 
it was pointed out in Ref. ~\cite{FCatm} that there is another 
interesting possibility of explaining the 
SK sub-GeV (SG) and multi-GeV(MG) atmospheric neutrino data 
by means of massless neutrino conversion $\nu_\mu \rightarrow \nu_\tau$ 
induced by FC as well as FD interactions 
with matter in the Earth. 
In this case the evolution equation for the system of 
$\nu_\mu-\nu_\tau$ is described by the same form as 
in (\ref{motion}) but without the standard matter 
potential $\sqrt{2}\,G_F n_e(r)$ in the Hamiltonian
matrix in eq. (\ref{hamil}).  

This proposal has been criticized in Ref.~\cite{FCatm2} where it was 
claimed that such FC induced neutrino oscillation solution
would be ruled out if one takes into account the upward going passing
muon (P$_\mu$) as well as the stopping muon (S$_\mu$) data. 

We have reexamined in Ref.~\cite{taup99} the FC/FD solution 
following the prescriptions described in Ref.~\cite{Fogli98}. 
By the method of a $\chi^2$ analysis we compare the expected 
number of SG and MG events as well as P$_\mu$ and S$_\mu$ 
fluxes obtained by our computations to that of the SK data, 
corresponding to 52 kTy for the SG and MG samples, 923 days 
for P$_\mu$ and 902 days for S$_\mu$ muons~\cite{nakahata99}. 

In Figs. 3 and 4, we show the best fitted zenith angle 
distributions for (SG, MG) and (P$_\mu$, S$_\mu$), respectively. 
From these plots we can see the the fit to the data is 
quite poor for the passing muon sample (see thick
curves in Fig. 4). 
%
%%%%%%%%%%%%%%%%%%%%%%%%%%%%%%%%%%%%%%%%%%%%%%%%%%
\begin{figure}
\vglue 0.3cm
\centering\leavevmode
\epsfxsize=205pt
\epsfbox{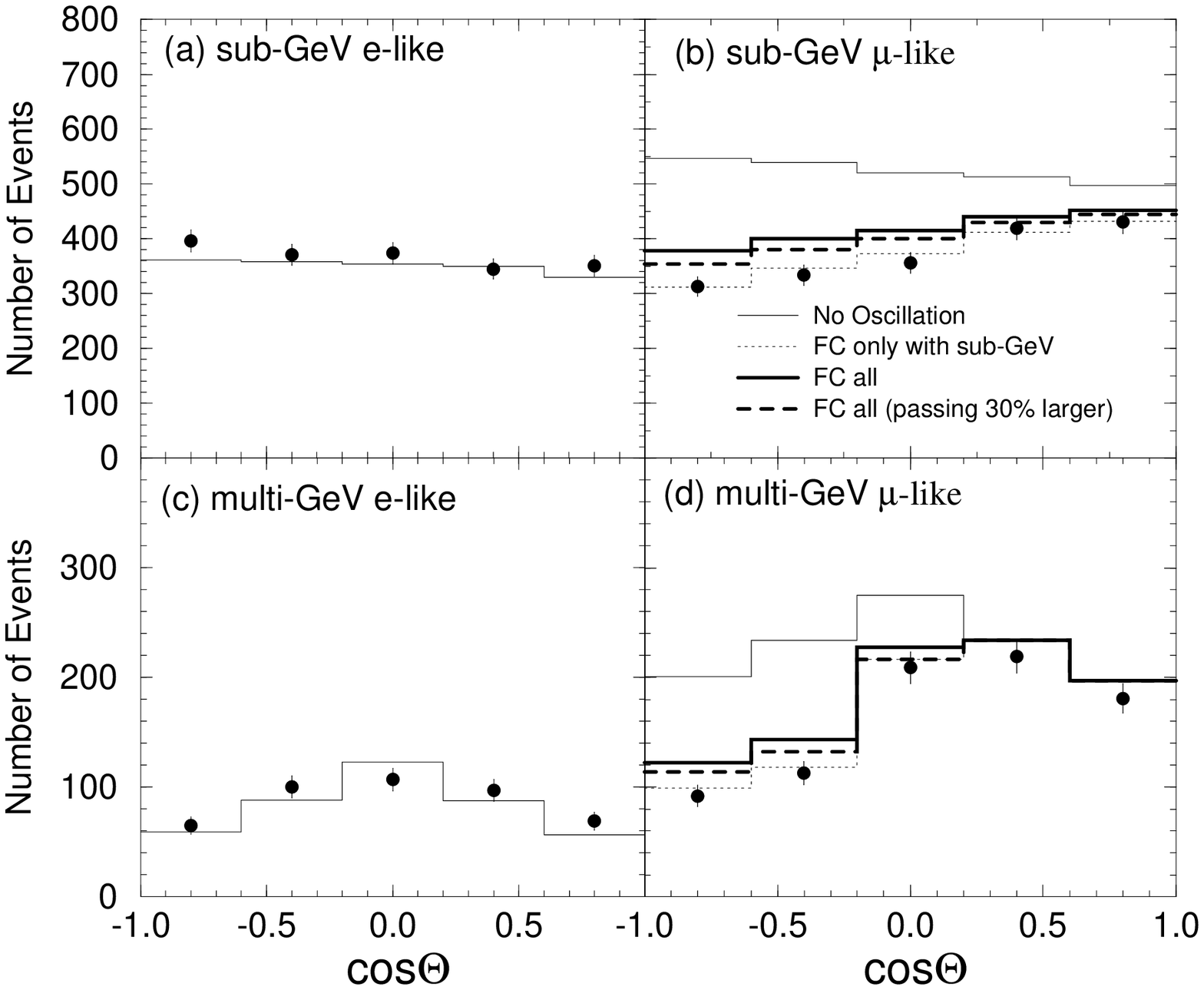}
\vglue -1.5cm
\caption{{\small 
Best fitted zenith angle distributions for SK SG and 
MG samples with the best fitted parameters. 
(Taken from Ref.~\cite{taup99}.)
}}
\vglue 0.25cm
\centering\leavevmode
\epsfxsize=195pt
\epsfbox{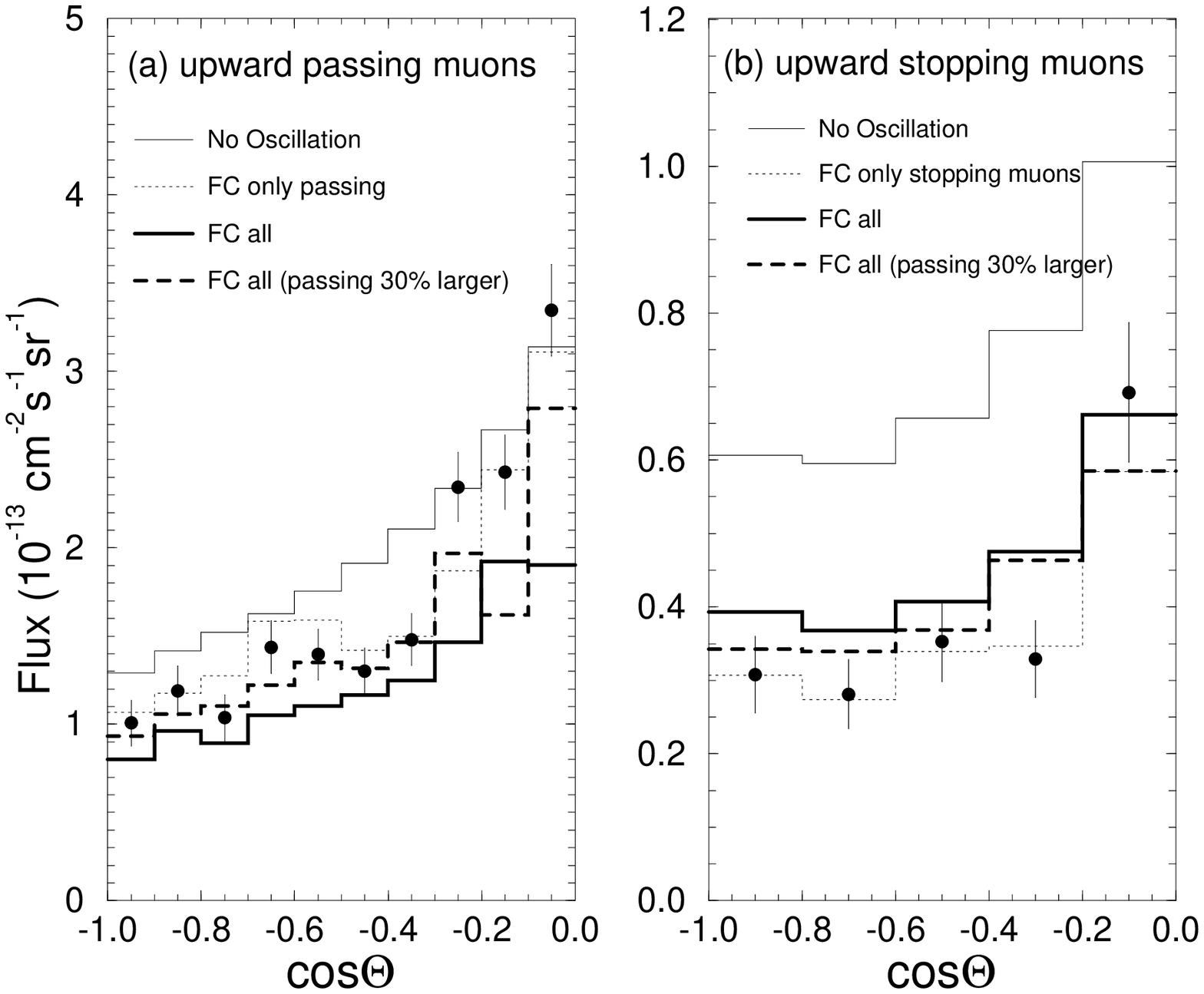}
\vglue -1.1cm
\caption{{\small
Same as in Fig. 4 but for P$_\mu$ and S$_\mu$ samples.
Best fitted zenith angle distributions for SK 
P$_\mu$ and S$_\mu$ with the best fitted parameters. 
of FC/FD conversion scenario. 
(Taken from Ref.~\cite{taup99}.)
}}
\vglue -0.7cm
\end{figure}
In order to compare quantitatively the quality of 
the fit of the FC/FD induced oscillation with the mass 
induced one, in Table 1 we show $\chi^2_{min}$ values and C.L. (in \%) 
for both scenarios, for individual SK samples as well 
as for several combinations.  
From this table we notice that the fit is very poor if 
P$_\mu$ sample is combined with any other SK samples. 
If we exclude P$_\mu$ from the fit FC/FD solution can be 
considered as good as the mass induced one.
While these results are based on somewhat old data, this conclusion 
does not seem to change even if we would have used the most recent 
SuperKamiokande data~\cite{skatm_ichep2000}. 
%

%
%%%%%%%%%%%%%%  Table I  %%%%%%%%%%%%%%%%%%%%%%%%%%%%%%%%%%%%%%%%%%%%%%%%%
%
\begin{table}[h]
\vglue -0.5cm
\caption[Tab]
{{\small 
$\chi^2_{\rm{min}}$ (C.L. in \%) values: FC/FD vs. Mass ind. osc.
(Taken from Ref.~\cite{taup99}.) }
}
\begin{center}
\begin{tabular}{c|c|c|c}
\hline
\hskip -0.5cm Sample  & FC/FD &  Mass   & d.o.f.   \\ 
\hline
\hskip -0.5cm SG  &   3.3(91)  & 3.2(92)   &  8   \\ 
\hskip -0.5cm MG  &   7.8(45)  & 7.6(47)  &  8   \\ 
\hskip -0.5cm P$_\mu$ &   12.9 (12) & 13.0(11)    &  8   \\ 
\hskip -0.5cm S$_\mu$ &   1.2 (75)& 1.6(66)   &  3   \\ 
\hline
\hskip -0.5cm SG+MG  &   11.2(89)  & 13.7(75)   &  18   \\ 
\hskip -0.5cm P$_\mu$ + S$_\mu$  &   41.2(0.01)  & 17.1(82)   &  13   \\ 
\hskip -0.5cm SG+MG+P$_\mu$ &   91.7($10^{-6}$)  & 32.8(24)  &  28   \\ 
\hskip -0.5cm SG+MG+S$_\mu$ &   13.7(94)  & 20.0(64)   &  23   \\ 
\hline
\hskip -0.5cm All &   109.3($10^{-8}$)  & 37.4(28)   &  33   \\ 
\hskip -0.2cm All\ (P$_\mu$ 30 \% $\nearrow$) &   54.9 (1) & ---   &  33   \\ 
\hline
\end{tabular}
\end{center}
\label{tab1}
\vskip -1.15cm
\end{table}
\vglue 0.3cm 

Our results described above can be qualitatively understood 
if we note the fact that observed P$_\mu$ is indicating 
rather weak deficit $\sim 15$ \% whereas all the 
other SK samples are indicating substantially 
stronger deficits $\sim 50$ \% of upward going 
neutrino fluxes, which is difficult to reconcile 
by any energy independent conversion mechanism. 

Moreover, as it was pointed out in Ref.~\cite{FCatm2}, 
the observed value of the double ratio of the 
passing over stopping sample $R(s/p) \equiv
(stop/through)_{data}/(stop/through)_{MC}$
$ = 0.63 \pm 0.1 $~\cite{nakahata99} is significantly 
smaller than unity which also disfavors FC/FD 
conversion scenario because it 
predicts $R(s/p)$ to be almost unity. 
It seems the only possibility to recover FC/FD induced oscillation 
as an acceptable solution is to assume a significantly larger ($> 30$ \%)  
P$_\mu$ flux than currently expected value but without
increasing the other lower energy neutrino fluxes. 

We note that several features of our results and observations 
are qualitatively similar to the ones obtained in 
Ref.~\cite{fc_val} but not quantitatively the same. 
Our results disfavor pure FC/FD oscillation solutions more 
strongly than Ref.~\cite{fc_val} but weaker than that in 
the first paper in Ref.~\cite{FCatm2}. 
This is essentially because the correlations of errors assumed 
in our $\chi^2$ analysis are substantially stronger than 
the ones used in Ref.~\cite{fc_val} though they are weaker than 
the ones considered in the first paper in Ref.~\cite{FCatm2}.

We also note that in Ref.~\cite{BGP}, magnitudes of 
$\epsilon$ and $\epsilon'$ parameters relevant for 
atmospheric neutrinos has been disfavored by the model 
independent analysis 
based on the negative results 
of the lepton flavor-violating decay or lepton 
universality violation searches.  
Finally, let us stress that the suggested FC/FD solution 
could be independently tested or excluded by the long-baseline 
neutrino oscillation experiments~\cite{fclbl}.

%%%%%%%%%%%%%%%%%%%%%%%%%%%%%%%%%%%%%%%%%%%%%%%%%%%%%%%%%%%%%%%%%%%%%%
%% Section IV %%%%%%%%%%%%%%%%%%%%%%%%%%%%%%%%%%%%%%%%%%%%%%%%%%%%%%%%
%%%%%%%%%%%%%%%%%%%%%%%%%%%%%%%%%%%%%%%%%%%%%%%%%%%%%%%%%%%%%%%%%%%%%%
\section{Neutrino Factory}

Finally, let us briefly discuss the possibility to test 
this kind of new interactions at future neutrino oscillation 
experiments at neutrino factory which use intense
neutrino beams from a muon storage ring~\cite{Geer}. 

First we consider the case where solar neutrino problem 
is solved by massless neutrino oscillation induced by 
FC as well as FD interactions. 
Since only $\nu_e \to \nu_\tau$ transitions are viable,
an independent test would require a $\nu_\tau$ ($\bar\nu_\tau$)
appearance experiment using an intense beam of $\nu_e$
($\bar{\nu}_e$), which could be created at future 
neutrino factories~\cite{Geer}.

Assuming a constant density and using the approximation that $n_d
\simeq n_u \simeq 3n_e$ in the earth, 
and the approximation $n_e \sim 2$ mol/cc (which is valid close 
to the earth surface), we find that~\cite{BGHKN} 
for the case of non-standard neutrino scattering off 
$d$-quark, $P_{e\tau} \equiv P(\nu_e\to\nu_\tau) 
\sim$ few $\times 10^{-4}$ 
for the K2K distance ($L = 250$~km) 
and $P_{e\tau}\sim$ few $\times 10^{-3}$ for the MINOS one ($L = 732$~km) 
for our best fit parameters. 
Similarly for $u$-quark, $P_{e\tau}\sim$ few $ \times 10^{-5}$
for the K2K distance and $P_{e\tau}\sim$ few $\times 10^{-4}$ for 
the MINOS one for the best fit parameters.  
These estimates imply that it would be hard but not
impossible, at least for the case of scattering off $d$-quarks, to
obtain some signal of $\nu_e \to \nu_\tau$ conversions due to FC/FD
interactions by using an intense $\nu_e$ beam which can be created by
a muon storage ring~\cite{Geer}.
Of course, it must be confirmed that the signal is not
due to the mass induced oscillation by studying for e.g. 
the energy dependence of the conversion probability. 

For atmospheric neutrinos, as we have discussed in sec. 3, 
it seems unlikely that FC/FD induced oscillation is the dominant
cause of the deficit since the fit is quite poor. 
However, to some extent FC/FD interactions can induce some 
subdominant effect in the usual mass induced ones
which would not lead to any contradiction 
with the observed data. 
Assuming that atmospheric neutrino data is mainly explained by 
the mass induced oscillation one can test the presence of such 
non-standard interactions by the future oscillation experiment
at neutrino factories~\cite{nufactfc}. 

It can be shown~\cite{nufactfc} that FC/FD interactions can 
induce some ``fake'' CP violating effect in matter even if
the CP violating phase is absent. 
Let use define $\Delta P(\nu_\mu \to \nu_{\tau})$ as 
$\Delta P(\nu_\mu \to \nu_{\tau})
\equiv P(\nu_\mu \to \nu_{\tau})
-P(\bar{\nu}_\mu \to \bar{\nu}_{\tau})$ where 
$\Delta P$ is in general functions of mass-mixing parameters, 
neutrino energy, the distance of the neutrino source to 
the detector as well as FC/FD parameters ($\epsilon$ and 
$\epsilon'$) we are assuming. 
We can say that if $\Delta P(\nu_\mu \to \nu_{\tau})$ is larger 
than the maximally expected value from the pure mass-induced 
oscillation with CP violating phase, this could be an 
indication of such FC/FD interactions or the presence 
of 4 (or more) neutrino mixing with sterile neutrino(s). 

\begin{figure}
\vglue 0.3cm
\hglue -0.4cm
\centering\leavevmode
\epsfxsize=220pt
\epsfbox{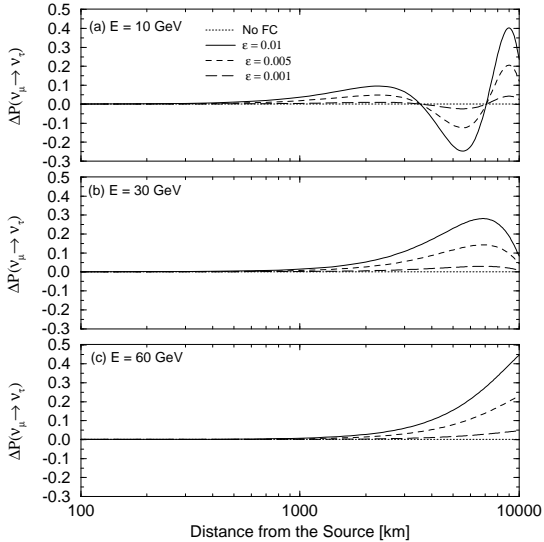}
\vglue -0.6cm
\caption{
{ \small
$\Delta P(\nu_\mu \to \nu_{\tau})
\equiv P(\nu_\mu \to \nu_{\tau})
-P(\bar{\nu}_\mu \to \bar{\nu}_{\tau})$ 
are plotted for various different neutrino energies and values of 
$\epsilon$ (we set $\epsilon'=0$) as a function of 
distance from the source. 
Here, FC interaction with $d$ (or $u$) quarks is assumed with 
the approximation $n_d=n_u=3n_e$ with  the constant electron 
number density $n_e = 2$ mol/cc. 
Here, also non-zero neutrino masses and mixing angles are assumed
only for $\nu_\mu$ and $\nu_\tau$ system with 
$\Delta m^2 = 3.5 \times 10^{-3}$ eV$^2$, and 
$\sin^2 2\theta = 1.0$ assuming oscillation solution 
to the atmospheric neutrino observation. 
} }
\vglue -0.8cm
\end{figure}

Based on this idea we have computed $\Delta P$ as a function 
of the distance from the source for various values of 
FC interaction parameter as well as neutrino energy. 
In Figs. 5 and 6 we plot our results for the case 
with pure 2 generation and 3 generation, respectively. 
From Fig. 5 we can see that even for 2 generation, 
$\Delta P$ can be non-zero  and can be relatively 
large for $L \gsim $ few 1000 km  
even for $\epsilon$ smaller than 1 \%. 
If $\nu_\mu$ is mixed only with $\nu_{\tau}$, 
$\Delta P \ne 0$ can be an indication of 
the presence of FC/FD interactions.  

If there are mixing in 3 generation and CP violating phase
things are more complicated since $\Delta P$ could be
different from zero either due to the standard matter effect 
or due to the CP violating phase or the combinations of both 
effects and we need to see if such effect, especially 
the standard matter effect (since the effect of CP phase
is expected to be small), is small enough in order 
to constrain FC/FD interactions. 

In deed, we can see from Fig. 6 that for higher energy, 
the standard matter effect as well as the CP violating effect 
from pure mass induced oscillation are suppressed compared with 
the case of lower energy.  
Therefore, at such higher energy, the FC/FD effects are
enhanced with respect the the standard matter and/or CP violating 
effect 
and $\Delta P \ne 0$ can be an indication of 
the presence of FC/FD interactions, or the absence of such 
effect can constrain these interactions. 
More detailed discussions will be presented in Ref.~\cite{nufactfc}.

\begin{figure}
\vglue 0.3cm
\hglue -0.4cm
\centering\leavevmode
\epsfxsize=220pt
\epsfbox{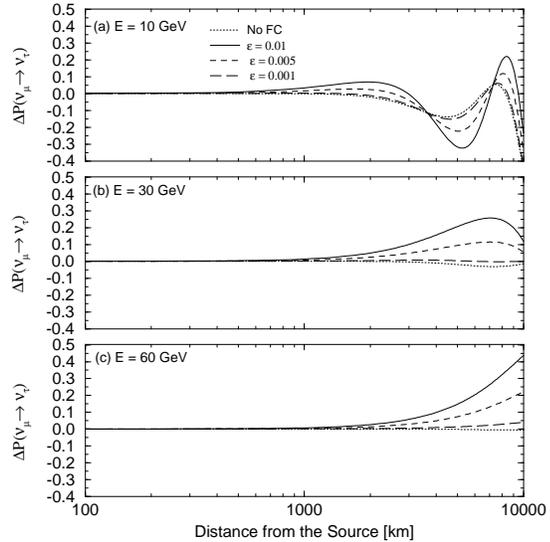}
\vglue -0.6cm
\caption{
{ \small 
Same as in Fig. 5 but for a three neutrino system. 
Here, we take 
$\Delta m^2_{12} =  5.0 \times 10^{-5}$ eV$^2$, 
$\sin^2 2\theta_{12} = 0.8$ assuming MSW LMA solution, 
$\Delta m^2_{13} \simeq \Delta m^2_{23} = 3.5 \times 10^{-3}$ eV$^2$, 
and 
$\sin^2 2\theta_{23} = 1.0$ assuming $\nu_\mu-\nu_\tau$ channel 
atmospheric neutrino solution, 
and $\sin^2 2\theta_{13} = 0.1$ and CP 
violating phase $\delta = \pi/2$. 
} }
\vglue -0.8cm
\end{figure}

%%%%%%%%%%%%%%%%%%%%%%%%%%%%%%%%%%%%%%%%%%%%%%%%%%%%%%%%%%%%%%%%%%%%%%
%% Section V %%%%%%%%%%%%%%%%%%%%%%%%%%%%%%%%%%%%%%%%%%%%%%%%%%%%%%%%
%%%%%%%%%%%%%%%%%%%%%%%%%%%%%%%%%%%%%%%%%%%%%%%%%%%%%%%%%%%%%%%%%%%%%%

\section{Conclusions}

Although the mass induced oscillation is the most plausible explanations
for the solar as well as atmospheric neutrino observations, 
it is interesting to consider some alternative solutions, which 
must eventually be excluded (or confirmed) by the experiments. 
We have studied if neutrino oscillations induced by 
FC as well as FD interactions can be such solutions. 
We found that while it is unlikely that atmospheric neutrinos
can be explained only by such interactions, 
solar neutrinos can be well accounted for by the 
massless neutrino oscillations induced by such 
non-standard interactions. 
We have also discussed that future neutrino oscillation 
experiments can test such kind of non-standard interactions. 

%%%%%%%%%%%%%%%%%%%%%%%%%%%%%%%%%%%%%%%%%%%%%%%%%%%%%%%%%%%%%%%%%%%%%%
%% Acknowledgments %%%%%%%%%%%%%%%%%%%%%%%%%%%%%%%%%%%%%%%%%%%%%%%%%%%
%%%%%%%%%%%%%%%%%%%%%%%%%%%%%%%%%%%%%%%%%%%%%%%%%%%%%%%%%%%%%%%%%%%%%%

\section*{Acknowledgments}

This work was supported by Funda\c{c}\~ao de Amparo \`a Pesquisa
do Estado de S\~ao Paulo (FAPESP) and by Conselho Nacional de e 
Desenvolvimento Cient\'\i fico e Tecnolog\'ogico (CNPq).
I thank S.\  Bergmann, A.\ M.\ Gago, M.\  M.\  Guzzo, 
P.\ C.\  de Holanda, P.\  I.\  Krastev, O.\ L.\ G.\ Peres and 
R.\ Zukanovich Funchal for discussions and collaborations. 

%%%%%%%%%%%%%%%%%%%%%%%%%%%%%%%%%%%%%%%%%%%%%%%%%%%%%%%%%%%%%%%%%%%%%%
%% References %%%%%%%%%%%%%%%%%%%%%%%%%%%%%%%%%%%%%%%%%%%%%%%%%%%%%%%%
%%%%%%%%%%%%%%%%%%%%%%%%%%%%%%%%%%%%%%%%%%%%%%%%%%%%%%%%%%%%%%%%%%%%%%

\end{document}